\documentclass[a4paper,12pt]{article}
 \pdfoutput=1
\usepackage{amssymb}
\usepackage{amsmath}
\usepackage{epsfig}
\usepackage{graphicx}
\usepackage{slashed}
\usepackage{xcolor,color}
\usepackage{ulem}
\usepackage{subfigure}
\usepackage[toc,page]{appendix}
\usepackage[makeroom]{cancel}
\usepackage{tikz}
\usetikzlibrary{shapes.geometric,positioning,decorations.pathreplacing,decorations.pathmorphing,patterns,decorations.markings}

\makeatletter

\@addtoreset{equation}{section}
\makeatother
\def\be{\begin{equation}}
\def\ee{\end{equation}}
\def\bea{\begin{eqnarray}}
\def\eea{\end{eqnarray}}
\def\bgn{\begin{align}}
\def\egn{\end{align}}

\def\({\left(}
\def\){\right)}
\def\<{\left<}
\def\>{\right>}

\def\({\left(}
\def\){\right)}
\def\<{\left<}
\def\>{\right>}
\def\!{\right|}
\def\|{\left|}

\def\[{\left[}
\def\]{\right]}

\def\+{\bar}

\def\i{{\bar{i}}}

\def\psit{|\psi(t)\rangle}

\usepackage{hyperref}
\hypersetup{colorlinks,%
  citecolor=blue,%
  linkcolor=blue,%
  pdftex}

\begin{document}

\begin{titlepage}
\vskip1cm
\begin{flushright}
\end{flushright}
\vskip0.25cm
\centerline{
\bf \large 
Machine Learns Quantum Complexity
} 
\vskip0.8cm \centerline{ \textsc{Dongsu Bak$^{a}$,
Su-Hyeong Kim$^{a,b}$, Sangnam Park$^{a}$,
Jeong-Pil Song$^{a}$
}
}
\vspace{0.8cm} 
\centerline{\sl  
a) Physics Department \& Natural Science Research Institute}
\centerline{\sl University of Seoul, 163 Siripdaero, Seoul 02504,  Korea}
 \vskip0.2cm
\centerline{\sl b) Department of Physics and Photon Science}
\centerline{\sl Gwangju Institute of Science and Technology,\\123 Cheomdan-gwagiro, Gwangju 61005, Korea}

 \vskip0.3cm

 \centerline{
\tt{\small (dsbak@uos.ac.kr,ksh46378@gm.gist.ac.kr,u98parksn@gmail.com,jeongpilsong@gmail.com)}
} 
  \vspace{1.5cm}
\centerline{ABSTRACT} \vspace{0.65cm} 
{
We study how a machine based on deep learning algorithms learns Krylov spread complexity in quantum systems with $N \times N$ random Hamiltonians drawn from the Gaussian unitary ensemble. Using thermofield double states as initial conditions, we demonstrate that a convolutional neural network-based algorithm successfully learns the Krylov spread complexity across all timescales, including the late-time plateaus where states appear nearly featureless and random. Performance strongly depends on the basis choice, performing well with the energy eigenbasis or the Krylov basis but failing in the original basis of the random Hamiltonian. The algorithm also effectively distinguishes temperature-dependent features of thermofield double states.
Furthermore, we show that the system time variable of state predicted by deep learning is an irrelevant quantity, reinforcing that the Krylov spread complexity well captures the essential features of the quantum state, even at late times.
}

\end{titlepage}



\section{Introduction}


Recently, the concept of quantum complexity \cite{Susskind:2014rva,Brown:2015lvg}, which quantifies the difficulty of preparing a given quantum state from a reference state using a specified set of simple (unitary) 
operations, has received considerable attention \cite{Brown:2015bva,Belin:2021bga,Iliesiu:2021ari,Caputa:2021sib,Brown:2021euk,Rabinovici:2020ryf,Zhai:2024tkz}.
In particular, the Krylov spread complexity, first introduced in \cite{Balasubramanian:2022tpr}, is a state-complexity measure that characterizes the spread of a quantum state during its Hamiltonian time evolution \cite{Nandy:2024evd}. Specifically, it quantifies how the state evolves and spreads within the Krylov subspace, which is spanned by repeated applications of the system's Hamiltonian on an initial state. 
(For the Krylov operator complexity, see  for instance \cite{Parker:2018yvk}.)
In a chaotic system, such as one described by a random Hamiltonian, the Krylov spread complexity exhibits a universal behavior over time \cite{Balasubramanian:2022tpr,Erdmenger:2023wjg,Cotler:2017jue,Hornedal:2022pkc}. This behavior includes a long ramp, a peak, and a subsequent slope that leads to a plateau. Notably, the ramp extends exponentially long \cite{Brandao:2019sgy,Haferkamp:2021uxo}, up to a timescale that is exponential in the entanglement entropy of the chaotic system ({\it e.g.}, $N \sim e^S$ in our case below). This universal behavior is closely related to that observed in random matrix theory (RMT) \cite{Guhr:1997ve,Bhattacharjee:2024yxj}, which is further argued to be connected to the late-time dynamics of black holes \cite{Cotler:2016fpe}.

In this note, we consider a RMT with an $N\times N$ Gaussian unitary ensemble (GUE) of matrices where a set of systems is specified by Hamiltonians randomly drawn from this ensemble. For our investigation of the Krylov spread complexity, we take the initial state to be a thermofield double (TFD) state for a given inverse temperature $\beta$ (details provided below), which is dual to a two-sided black hole in the context of the AdS/CFT correspondence \cite{Maldacena:2001kr}. Using time-series data of the complexity along with the corresponding $N$-dimensional quantum states, we aim to investigate how a machine learns the complexity of a state at a given moment in time, particularly as the structures of the states become increasingly complicated and random. 

For this purpose, we primarily 
employ deep learning (DL) method \cite{LeCun:2015pmr} based on convolutional neural network (CNN) algorithms \cite{Lecun:1998iof}. At first glance, one might naively expect that DL would work effectively only in the early-time region, where the structural patterns of states are more pronounced. Interestingly, however, our DL algorithm performs well even in the late-time region ($t \gtrsim  N$), including the plateau where the states become almost featureless and random.

To test the DL algorithm further, we examine the basis dependence of its efficiency, as any physical information should be independent of the choice of basis. We find that the algorithm's performance depends dramatically on the chosen basis. It performs well in the energy eigenbasis or the Krylov basis, while the performance is extremely poor in the original basis, \textit{i.e.}, the basis used prior to diagonalization into the energy eigenbasis.

We also consider training data that includes TFD states at different temperatures. In this case, even the complexity values in the plateau region are crucially dependent on the temperature of the TFD states. Our results show that the DL algorithm effectively differentiates key features across multiple temperatures without any issues.

Finally, we test the algorithm with respect to other variables, such as the system time of the quantum states. These tests suggest that the time variable is physically irrelevant for characterizing the state, particularly in the late-time region.

This paper is organized as follows. In Section \ref{sec2}, we briefly review the Krylov spread complexity. We also introduce the TFD initial state and the RMT with the GUE to specify the system with the corresponding random Hamiltonian. In Section \ref{sec3}, we review our DL method, focusing on the CNN algorithm. In Section \ref{sec4}, we present the main results of this study. The final section is devoted to our conclusions and discussions.

\section{Krylov Spread Complexity}\label{sec2}

In this section, we shall briefly review the basic construct of the Krylov spread complexity, which was first introduced in \cite{Balasubramanian:2022tpr}. Let us start with a quantum system with a Hamiltonian $H$ and the corresponding Hilbert space ${\cal H}$.
We are primarily interested in the spread of a time-evolving quantum state 
\be
\label{Eq_HeisenbergState}
  \psit = e^{-iHt}|\psi(0)\rangle = \sum_{n} \frac{(-it)^n}{n!} H^n|\psi (0)\rangle
\ee 
 in the Hilbert space, starting from an initial state $|\psi (0)\rangle$. In order to measure the Krylov spread complexity, one first introduces the so-called Krylov basis $\mathcal{K}=\{|K_n\rangle \mid n=0,1,2,\dots\}$, which may be constructed by the recursive Gram-Schmidt procedure, called the 
Lanczos method \cite{Lanczos:1950zz}, to the non-orthogonal basis $\{H^n|\psi(0)\rangle \mid n=0,1,2,\dots\}$. (This Krylov basis forms in general the Krylov subspace of the Hamiltonian.) Specifically this orthonormal Krylov basis may be obtained by the recursion relations
\begin{equation}\label{Eq_Lanczosalgorithm}
\begin{aligned}
  &|A_{n+1}\rangle = (H-a_n)|K_n\rangle - b_n|K_{n-1}\rangle, \\ 
  &|K_n\rangle = b_n^{-1}|A_n\rangle, \quad \ \ \ \,\,  |K_0\rangle \equiv |\psi(0)\rangle.
\end{aligned}
\end{equation}
Here the 
Lanczos coefficients $a_n$ and $b_n$, which basically contain all the dynamical information of the system, are defined by
\begin{align}\label{Eq_Lanczoscoeficients}
a_n = \langle K_n|H|K_n\rangle, 
    \quad  \  b_n = \langle A_n | A_n \rangle^{1/2}, 
\end{align}
with $b_0=0$. As a measure of the spread of the time-evolved wavefunction, the Krylov spread complexity at time $t$ is defined by
\begin{align}\label{Eq_Complexity}
    C(t) = 
 \sum_{n} n\,  |\langle K_n| \psi(t)\rangle|^2\,,
\end{align}
where the weight $n$ may be replaced by any positive increasing sequence of real numbers. In this definition, 
any other ordered orthonormal basis ${\cal B}=\{|B_n\rangle \mid n=0,1,2,\dots\}$ could be used  
instead of the Krylov basis ${\cal K}$. However it has been shown 
\cite{Balasubramanian:2022tpr} that the Krylov basis minimizes the complexity (\ref{Eq_Complexity})
over all possible 
choices of such bases $\cal B$. Note that, in this Krylov basis, the Hamiltonian becomes  tri-diagonal
as
\begin{align}
   H|K_n\rangle = b_n|K_{n-1}\rangle + a_n|K_n\rangle + b_{n+1}|K_{n+1}\rangle,
\end{align}
which is  called the
Hessenberg form of Hamiltonian. Furthermore, in this basis, the  Schr\"odinger 
equation becomes 
\begin{align}\label{Eq_amplitudeSchrodinger}
    i\dot{\psi}^K_n(t) = b_n\psi^K_{n-1}(t) + a_n\psi^K_{n}(t) + b_{n+1}\psi^K_{n+1}(t),
\end{align}
where $\psi^K_n(t)= \langle K_n| \psi(t)\rangle$. 
Various methods for computing the Lanczos coefficients are discussed in \cite{Balasubramanian:2022tpr}.
Since, in this note, we are only concerned with the finite-dimensional RMT \cite{Cotler:2017jue},  one can directly compute the Hessenberg form of Hamiltonian numerically, from which the corresponding Lanczos coefficients and the Krylov amplitudes can also be read directly. 

Below we shall take our initial state as a so-called TFD state in an inverse temperature $\beta=1/T$. 
We specialize in the energy eigenbasis of $H$ spanned by an energy eigenstate $|n\rangle$ with the corresponding energy eigenvalue $E_n$. The TFD state
\begin{align}\label{Eq_initialstate}
|\psi(0)\rangle = \frac{1}{\sqrt{Z}}\sum_{n} e^{-\frac{\beta}{2}E_n}|n,n\rangle,
\end{align}
purifies the thermal ensemble where 
$|n,n\rangle= |n\rangle_L \otimes |n\rangle_R $ belongs to the left-right doubled Hilbert space
${\cal H}\otimes {\cal H}$ and $Z =\sum_n e^{-\beta E_n}$ is the canonical partition function. With $H_{R}=1\otimes H$ and $H_L=H\otimes 1$, the TFD state is invariant under the boost transformation with a generator $(H_R-H_L)/2$. Hence the $H_L$ operation may be equivalently replaced by the
$H_R$ operation and  the general time evolution of the TFD system may be simply given by
$e^{-i H_R t}$ with  $e^{-i H_R t}|n,n\rangle=e^{-i E_n t}|n,n\rangle$. The left-right system in the TFD state is maximally entangled for a given temperature and dual to the two-sided black hole in the AdS/CFT correspondence \cite{Maldacena:2001kr}. It has been observed \cite{Balasubramanian:2022tpr,Erdmenger:2023wjg} that, especially for a chaotic system, there appears a universal behavior of the Krylov spread complexity which exhibits
a long ramp, a peak, a subsequent slope leading to a plateau as a function of time. In particular it shows an exponentially long ramp up to the time scale which is exponential in the entanglement entropy of the left-right system for a chaotic Hamiltonian. This universal behavior is closely related to that of the RMT \cite{Cotler:2017jue}  which is also closely related to the late-time behavior of the black hole dynamics \cite{Cotler:2016fpe}. In this note, we shall focus on a finite dimensional RMT with the GUE as a typical example of the Krylov spread complexity of the corresponding TFD state with a chaotic spectrum.

\begin{figure}[t] 
  \centering
  \includegraphics[width=12cm]{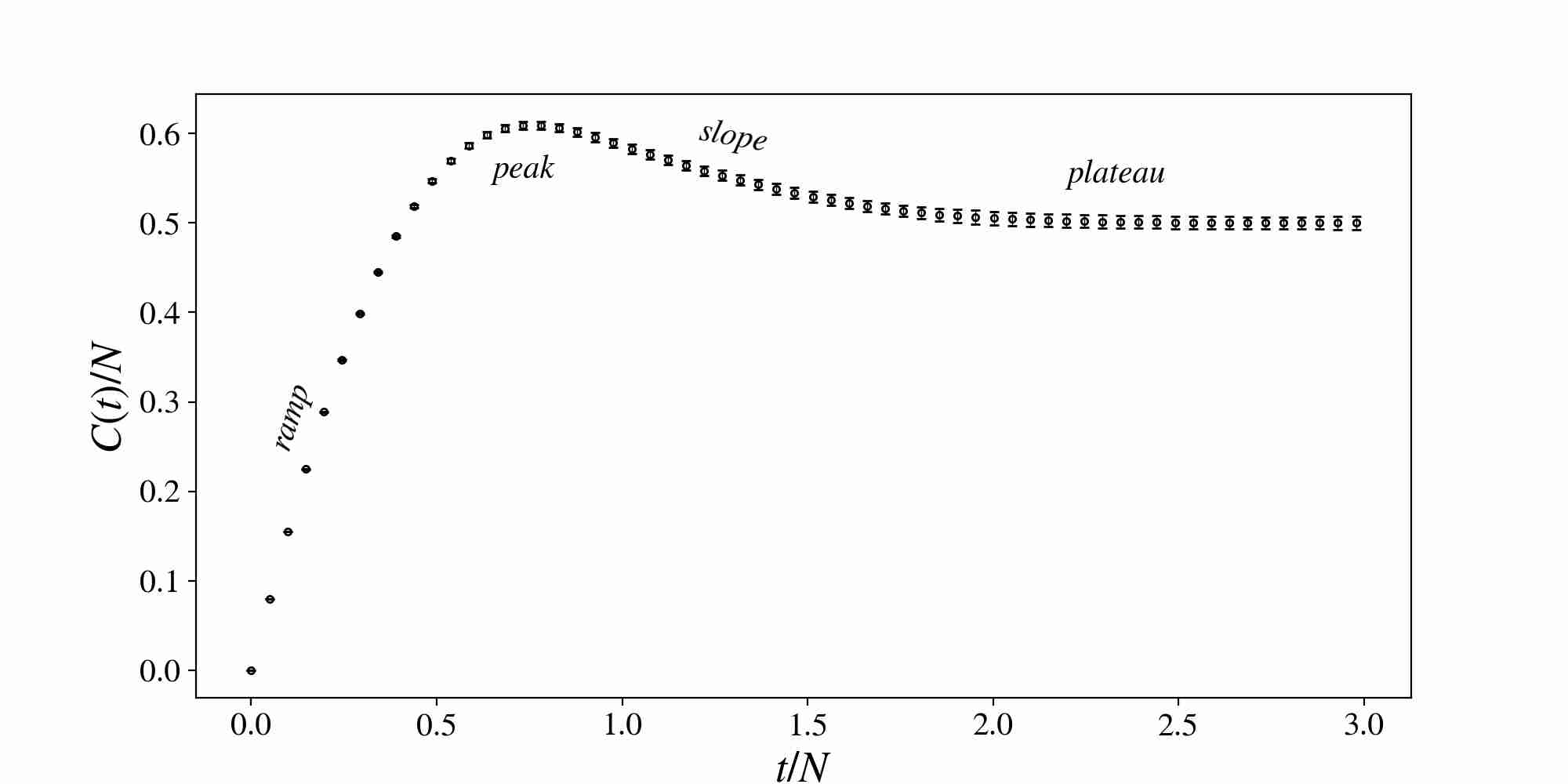} 
  \caption{\small Krylov spread complexity $C(t)/N$ for $N\times N$ random Hermitian matrices $H$ as a function of time $t/N$ with a $\beta=0$ TFD initial state and $N=1024$. 
As indicated on the plot we distinguish four phases: a ramp, a peak, a slope, and a plateau. The error bar denotes the standard deviation of samples.}
  \label{Fig_comp_GUE}
\end{figure} 

We calculate the Krylov spread complexity for $N\times N$ random Hermitian matrices $H$ in order to statistically analyze some important properties of the eigenvalues in the GUE statistics.
Here the Hamiltonian $H$ of degree $N$ is given as $H=(M+M^{\dagger})/2$ with complex elements $M_{ij}=u_{ij}+iv_{ij}$, where $u_{ij}$ and $v_{ij}$ are real numbers.
Note that the real entries $u_{ij}$ and $v_{ij}$ exhibit a normal distribution with zero mean ({\it i.e.}, $\langle u_{ij} \rangle = \langle v_{ij} \rangle =0$) and the variance of $1/N$.
In order to eliminate potential biases, we generated 200 different matrices $H$ to form random samples of size $N$ from the GUE distribution.
We have chosen the initial configuration at $t=0$ to be the TFD state.
In Figure~\ref{Fig_comp_GUE}, we present a benchmark study of the Krylov spread complexity $C(t)/N$ against time $t/N$ for the infinite temperature limit ($\beta=0$) and $N=1024$.
A key structure of the Krylov spread complexity, 
shown in Figure~\ref{Fig_comp_GUE}, may be largely characterized by four distinct regions, a ramp, a peak, a slope, and a plateau.
$C(t)$ rapidly increases in the small $t$ region of ramp until it reaches the peak.
In the region of slope a monotonous decrease in $C(t)$ was also found to saturate to a constant value at sufficiently large $t$. In the plateau region $C(t)$ remains a constant as a function of $t$ ({\it i.e.}, $dC(t)/dt \sim 0$). 



\subsection{Basis dependence on deep learning}\label{sec2.1}
A quantum state $\psit$ can be represented by different bases and of course they are connected by unitary transformations. Although all the physical information of a given Hamiltonian system should be completely independent of the basis, the DL performance seems to
be strongly dependent upon the choice of basis as will be shown below. In the following we shall study how a machine learns the Krylov spread complexity of quantum states 
directly from the data set of time-dependent states where we shall also be interested in the basis dependence of the DL performances. As mentioned earlier, we use the $N \times N$ random GUE Hamiltonian throughout this note. For the DL purposes, we generate $M$ different random Hamiltonian samples 
labeled by $s=1,2,\cdots, M$ and the corresponding Hamiltonian and eigenvalues are denoted by $H_{(s)}$ and $E^{(s)}_n\, (n=0,1,\cdots, N\negthinspace-\negthinspace1)$ respectively.  For simplicity, we denote the state   $|n,n\rangle_{(s)}$ by  $|{\cal E}^{(s)}_{n}\rangle$, explicitly  specifying the sample dependence. This notation is sensible because under the time evolution the TFD state remains within the subspace
spanned by $\{|{\cal E}^{(s)}_{n}\rangle \mid n=0,1,\cdots,N\negthinspace-\negthinspace1\}$. In this eigenbasis, the time evolution of the TFD state is represented by
\begin{align} \label{Eq_EigenState} 
& \psit_{(s)} 
=
 \frac{1}{\sqrt{Z_{(s)}}} \sum_{n} e^{-iE^{(s)}_n(t-\frac{i\beta}{2})}|{\cal E}^{(s)}_n\rangle= \sum_{n} \psi^{(s)}_n(t)|{\cal E}^{(s)}_n\rangle.
\end{align}
Using the Krylov basis constructed with the Gram-Schmidt procedure, the state can  also be expanded as
\begin{align}\label{Eq_KrylovState}
  \psit_{(s)} = \sum_{n} \psi_n^{K_{(s)}}(t)\,|K^{(s)}_n\rangle.
\end{align}
These two bases are naturally arising when we consider the Krylov spread complexity of the time evolving TFD state. One may create another basis using the unitary transformation 
\be 
  |\alpha_n\rangle_{(s)} = \sum_m
  U_{nm}^{(s)} |m\rangle_{(s)}
\ee
that brings back to the original basis before the diagonalization of each random Hamiltonian 
$H_{(s)}$ where $U_{(s)}$ is the corresponding unitary matrix and $|\alpha_n\rangle_{(s)}$ is 
the original basis before the diagonalization. We then define the new basis $ |R^{(s)}_n\rangle $ by $ |R^{(s)}_n\rangle = \sum_m U_{nm}^{(s)}\,|{\cal E}^{(s)}_m\rangle$ and then the state 
becomes 
\begin{align}\label{Eq_RandomState}
  |\psi(t)\rangle_{(s)} = \sum_{n} \psi_n^{R_{(s)}}(t)|R_n^{(s)}\rangle
\end{align}
with $\psi_n^{R_{(s)}}(t)=\sum_m (U_{(s)})^\dagger_{nm} \psi^{(s)}_m(t)$. From the sample-dependent 
unitary transformation, the amplitudes $\psi_n^{R_{(s)}}(t)$ look completely random while retaining all the physical information of the system. The last basis we consider is the pseudo-random basis that is produced by one particular sample-independent random unitary transformation, {\it e.g.} $U_{(0)}$. One may view that a particular type of ``noise'' is added to
the TFD amplitudes in the energy eigenbasis, which produces the new amplitudes in the pseudo-random basis.

\begin{figure}[t] 
  \centering
  \includegraphics[width=15.5cm]{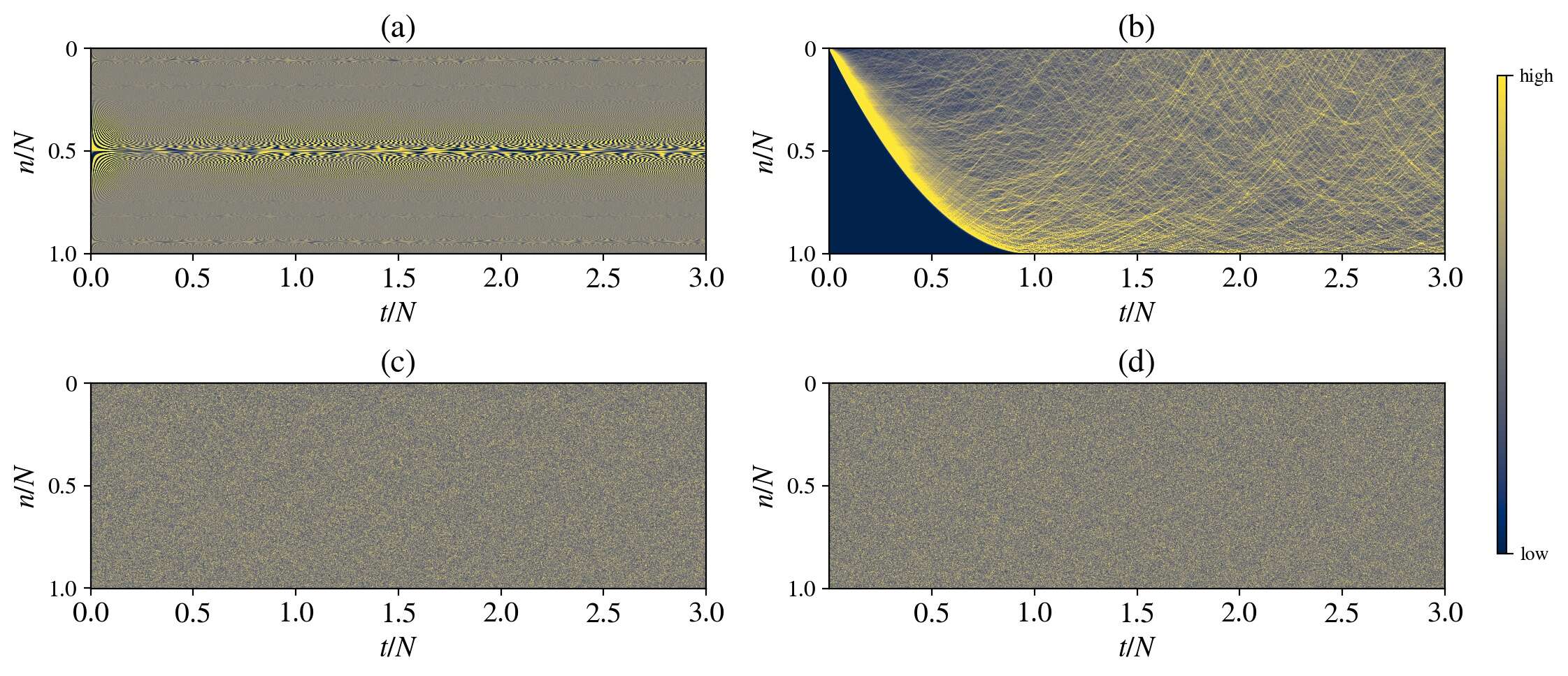} 
  \caption{\small 
  The magnitudes of the time-dependent amplitudes $|\text{Re}(\psi^{\cal B}_n(t)) + \text{Im}(\psi^{\cal B}_n(t))|$ for the intervals $0 \le n/N \le 1$ and $0 \le t/N \le 3$.
Here we considered four different choices of the basis, (a) the energy, (b) the Krylov, (c) the original, and (d) the pseudo-random basis.
}
  \label{Fig_state}
\end{figure} 

Figure~\ref{Fig_state} compares the magnitudes  of the combination
$|\text{Re}(\psi^{\cal B}_n(t)) + \text{Im}(\psi^{\cal B}_n(t))|$ for a single sample in the time region of $0 \le t/N \le 3$ for $0 \le n/N \le 1$. A remarkable feature of the states
for each basis is probably their unique fundamental structure. The semi-periodic pattern with a certain chaotic feature, shown in Figure~\ref{Fig_state}{\color{blue}(a)}, is quite noticeable in the choice of the energy eigenbasis. 
Vortex-like patterns emerge  as time progresses. 
As can be seen in Figure~\ref{Fig_state}{\color{blue}(b)}, the borderline where the state moves is apparent in the choice of the Krylov basis. Since the state moves along the 1D Krylov chain, it exhibits a zero-amplitude regime beyond the borderline at early times. The value of the amplitude is concentrated in the highest non-zero chain in the very early time; in the late time it diffuses to the lower level chain in Figure~\ref{Fig_state}{\color{blue}(b)}. While structured patterns are observed in the state evolution of the energy eigenbasis and Krylov bases, only disordered fluctuations are seen in the original and pseudo-random bases (See Figures~\ref{Fig_state}{\color{blue}(c)} and 
 {\color{blue}(d)}).

\section{Setup for Deep Learning}\label{sec3}
In this section, we will 
briefly review 
the DL 
 (a subset of machine learning) method  \cite{LeCun:2015pmr}  used in our study, focusing on two representative architectures of CNNs \cite{Lecun:1998iof} and fully connected networks (FCNs).
 DL models are characterized by their ability to automatically learn features 
 directly from raw input data, eliminating the need for manual feature engineering, which often fails to capture critical information. The features learned through this automated  
process enable DL models to identify complex patterns in data, making them highly effective in domains such as image recognition and natural language processing.


\subsection{Model Architecture}
DL methods basically use artificial neural networks, which consist of several layers of nodes (or neurons). Each node performs a simple operation on the input it receives from the nodes in the previous layer and then passes the result to the next layer.
 The following 
expression is a typical example of an FCN consisting of total $L$ layers of nodes including input layer,
\begin{equation}\label{FCN}
\begin{aligned}
 o_i\, &= f_{out}\left(\sum_{j=1}^{N_{(L-2)}} W^{(L-1)}_{ij} h^{(L-2)}_j + b^{(L-1)}_i\right), \\ 
  h^{(l)}_j &= f_h\left(\sum_{k=1}^{N_{(l-1)}} W^{(l)}_{jk} h^{(l-1)}_k + b^{(l)}_j\right),
\end{aligned}
\end{equation}
 where $o_i$ is the $i$-th component of the output in the final layer $L-1$, which has $N_{(L-1)}$ components. 
Here $l$ is the layer index, ranging from 1 to $L-1$. $h_j^{(l)}$ is the value of node $j$ in layer $l$, which has $N_{(l)}$ nodes. 
The input layer nodes are represented by $h_n^{(0)}=V_n$, the $n$-th component of the $N_{(0)}$ input variables.
$f_{out}$ is the activation function for the output nodes. This function (for example, $Softmax$, $Sigmoid$, or $Linear$) is selected based on the specific problem of interest.
Similarly the function $f_h$ represents the activation for the nodes $h^{(l)}$ in inner (or hidden) layers $0<l<L-1$, where usually a nonlinear function such as $tanh$ or Rectified Linear Unit (ReLU) is chosen for the activation. 
These nonlinear functions are used to introduce nonlinearity into the model, which is essential for learning complex patterns. Specifically, the function ReLU is defined as $f_{\text{ReLU}}(x)=\alpha x$ if $x>0$, otherwise $f_{\text{ReLU}}(x)=0$, with a typical choice of $\alpha>0$. The connection parameters $W$ and the bias $b$, which are 
randomly initialized in an appropriate manner \cite{Glorot:2010AISTATS}, represent the weights 
applied to the output vector of the nodes in the previous layer and the corresponding offset. These parameters 
will be repeatedly updated and optimized during training iteration by minimizing the loss associated with the task at hand.

The architecture of CNNs is derived by reorganizing the nodes of the hidden layers in FCNs. This involves narrowing the receptive fields of the connection weights and enabling weight sharing. In  weight sharing, nodes within parts (or channels) of the same layer have the same values and connection structures to nodes in the previous layer, ultimately performing convolutional operations.
The resulting architecture of the CNN example is described by the following 
expression, which applies to 2D image processing,
\begin{align}\label{CNN}
 h_{i,n_1,n_2}^{(l)}&= f_h\negthinspace\left(\sum_{j=1}^{N^{(l\negthinspace-\negthinspace1)}_{ch}}\negthinspace\sum_{m_1=0}^{K\negthinspace-\negthinspace1}\sum_{m_2=0}^{K\negthinspace-\negthinspace1} W_{i,j,m_1,m_2}^{(l)} h_{j,n_1+m_1,n_2+m_2}^{(l-1)}+b^{(l)}_i\negthinspace\right)\negthinspace\negthinspace, 
\end{align}
where $N_{ch}^{(l-1)}$ is the number of channels (kernels)
of layer $l-1$, $n_1\,(=1,\cdots,X_{(l)})$ and $n_2\,(=1,\cdots,Y_{(l)})$ represent the 2D pixel coordinates on the feature maps with sizes of ($X_{(l)}, Y_{(l)}$) in layer $l$. In the case of the input layer, $X_{(0)}$ and $Y_{(0)}$ correspond to the number of pixels along $x$ and $y$ axis of the input image. Assuming no padding is used\footnote{Padding refers to the process of adding extra pixels, typically zeros, around the borders of an image (or feature map) to control the spatial size reduction caused by convolution operations. In `valid' padding (i.e. no padding), the size of the feature map shrinks, while `same' padding maintains the input size by adding enough pixels to the borders.}, the feature map size at layer $l$ follows from the relations $X_{(l)} = X_{(l-1)}-K+1$ and $Y_{(l)} = Y_{(l-1)}-K+1$.
$m_1$ and $m_2$ representing convolution indices take integer values ranging from $0$ to $K-1$ with the kernel size $K$. Finally, the indices $i$ and $j$ represent the channel index of layer $l$ and $l-1$ respectively. 

CNNs are particularly effective for data with spatial and temporal hierarchies, such as images or time-series data, due to their ability to capture local dependencies using convolution operations. These operations scan the data with filters (kernels) to extract local patterns. After the convolution, pooling layers are typically applied to reduce the dimensionality of the data. Pooling is a process of 
selecting representative values (such as the maximum or average) from small regions within the feature map, effectively down-sampling the feature maps. This helps to preserve important features while making the model more computationally efficient and less prone to overfitting.
In contrast, FCNs do not inherently account for spatial structure but can capture global relationships across the entire input data set. This makes them more general but potentially less efficient for tasks where spatial structure is important. 

\begin{figure}[t] 
  \centering
  \includegraphics[width=13cm]{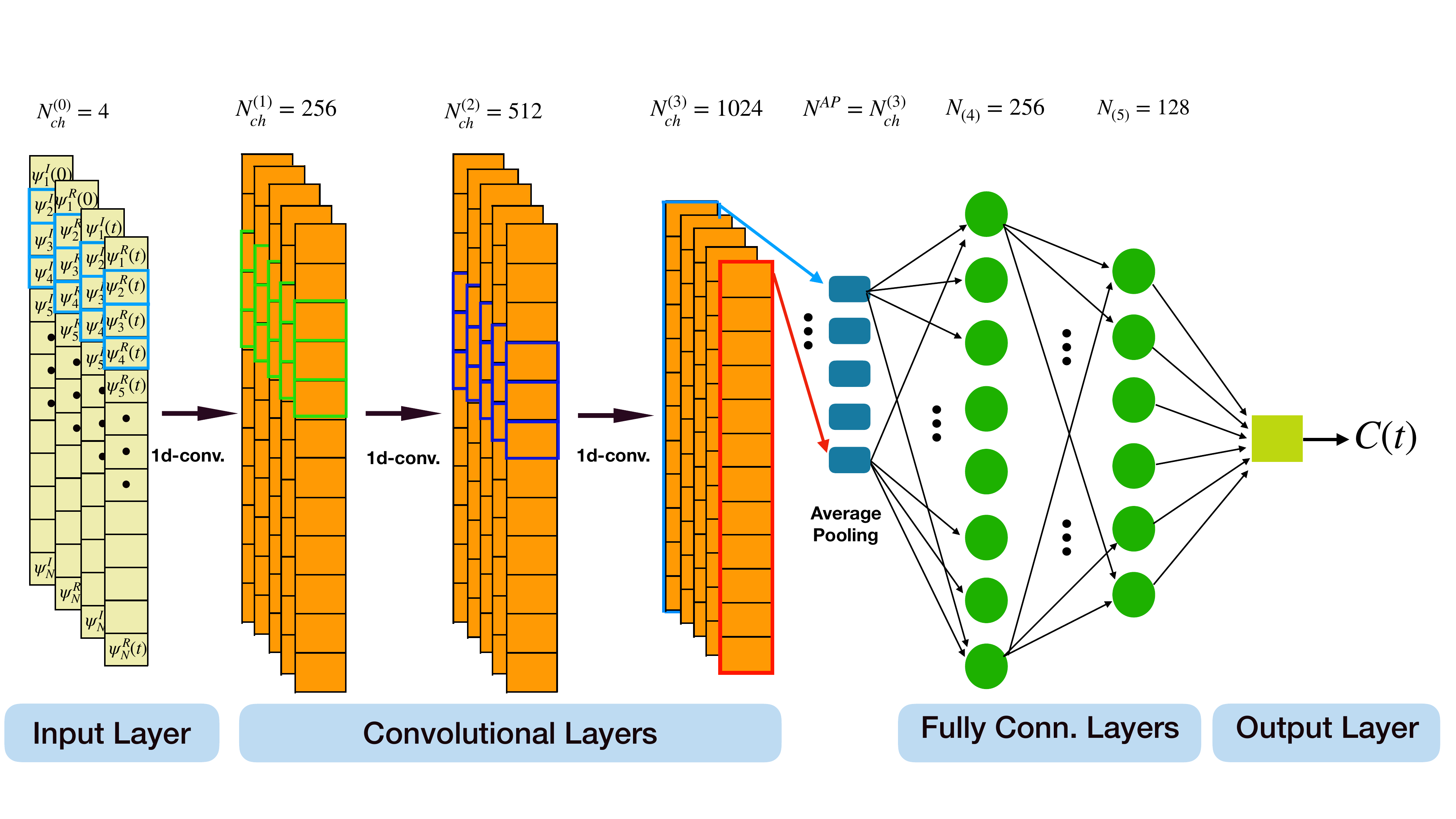}
  \caption{\small Our 1D CNN architecture. Starting from the left, this schematic diagram shows the input layer with 4 channels, representing the real and imaginary components of the $N =1024$ dimensional state vectors $|\psi(0)\rangle$ and $|\psi(t)\rangle$. Next are three convolutional layers with 256, 512, and 1024 channels, respectively, followed by an AveragePooling layer with $N^{(3)}_{ch}=1024$ pooling nodes ($h^{AP}_k=\frac{1}{X_{(3)}}\sum_{n=1}^{X_{(3)}}h^{(3)}_{k,n},k=1,\dots,N^{(3)}_{ch}$ : the red and light blue outlines in the third convolutional layer indicate the averaging over the entire region of the channel.). This is followed by two fully connected layers with 256 and 128 nodes, respectively, and finally, an output layer consisting of a single node. The light blue, light green, and navy blue elements displayed as three vertically stacked blocks in the input layer and the first two convolutional layers represent convolution kernels (e.g., in the first convolutional layer, $W_{i,j,m}^{(1)},\, i=1,...,256,\,j=1,2,3,4,\,m=0,1,2$) with   $K=3$ where the kernel size $K=3$ is chosen only for the sake of illustration.}
  \label{fig1DCNN}
\end{figure}

The 1D CNN 
in Figure~\ref{fig1DCNN} is our main architecture in this note, 
used to process four channels of 1024-dimensional vectors composed of real and imaginary components of both the initial state $|\psi(0)\rangle$ and  the current state $|\psi(t)\rangle$,  
\begin{equation}\label{1DCNN}
\begin{aligned}
  h_{i,n}^{(l)} &= f_h\left(\sum_{j=1}^{N^{(l-1)}_{ch}}\sum_{m=0}^{K-1} W_{i,j,m}^{(l)} h_{j,n+m}^{(l-1)} + b_i^{(l)}\right),\\
  (h_{1,n}^{(0)},h_{2,n}^{(0)},h_{3,n}^{(0)},h_{4,n}^{(0)}) &=  (Re[\psi_n(t)],Im[\psi_n(t)],Re[\psi_n(0)],Im[\psi_n(0)]),
\end{aligned}
\end{equation}
where the index $n$ runs from 1 to  $X_{(l)}$which denotes the size of feature arrays in layer $l$. (We include the initial state in the input variables as the complexity is defined with respect to the initial state.) Starting with an input vector size of $N = 1024$, the feature array sizes of convolutional layers are contracted by the relation $X_{(l)} = X_{(l-1)}-K+1$ across the layers as mentioned previously.  
The CNN architecture used here consists of three convolutional layers with progressively larger number of kernels ($N^{(1)}_{ch}=256$, $N^{(2)}_{ch}=512$, and $N^{(3)}_{ch}=1024$), allowing the model to capture both fine and coarse grained patterns of the data at the same time. 
The CNN layers are followed by a GlobalAveragePooling1D layer, which aggregates the feature arrays by averaging values of all nodes within the same channels as $h^{AP}_k=\frac{1\,\,}{X_{(3)}}\sum_{j=1}^{X_{(3)}}h^{(3)}_{kj}$, where $k=1,\dots, N^{(3)}_{ch}$. After the pooling layer, the data is passed through two fully connected layers $l=4$ and $l=5$ with $N_{(4)}=256$ and $N_{(5)}=128$ nodes, respectively. Both of these layers use the ReLU activation function, the same as the CNN layers. Finally, the output layer consists of a single node with a linear activation function $f_{out}(x)=x$, designed to output the estimated quantum complexity as a real value.

We experimented with both 1D CNNs using various kernel sizes and FCNs with different combinations of layers and nodes. We found that 1D CNNs mostly outperformed FCNs, with the best optimization achieved using a kernel size 
$K=5$. However, 
for the pseudo-random basis,
FCN appears to show slightly better 
performances than 
CNN 
 as will be illustrated later on. Having the same structure of the output layer of the CNN architecture, the FCN architecture used for the pseudo-random basis consists of 4-hidden layers with 1024, 512, 256, and 128 nodes, respectively. Each of these layers employs the ReLU activation function, the same as in the CNN architecture. In fact, the pseudo-random basis will not be our primary concern in this note as it is rather artificial and introduced for the sake of comparisons only. Hence we shall mainly use the 1D CNN model with $K=5$ as our primary architecture in our training of the complexity.

\subsection{Training}

\begin{figure}[t] 
  \centering
  \includegraphics[width=12.5cm]{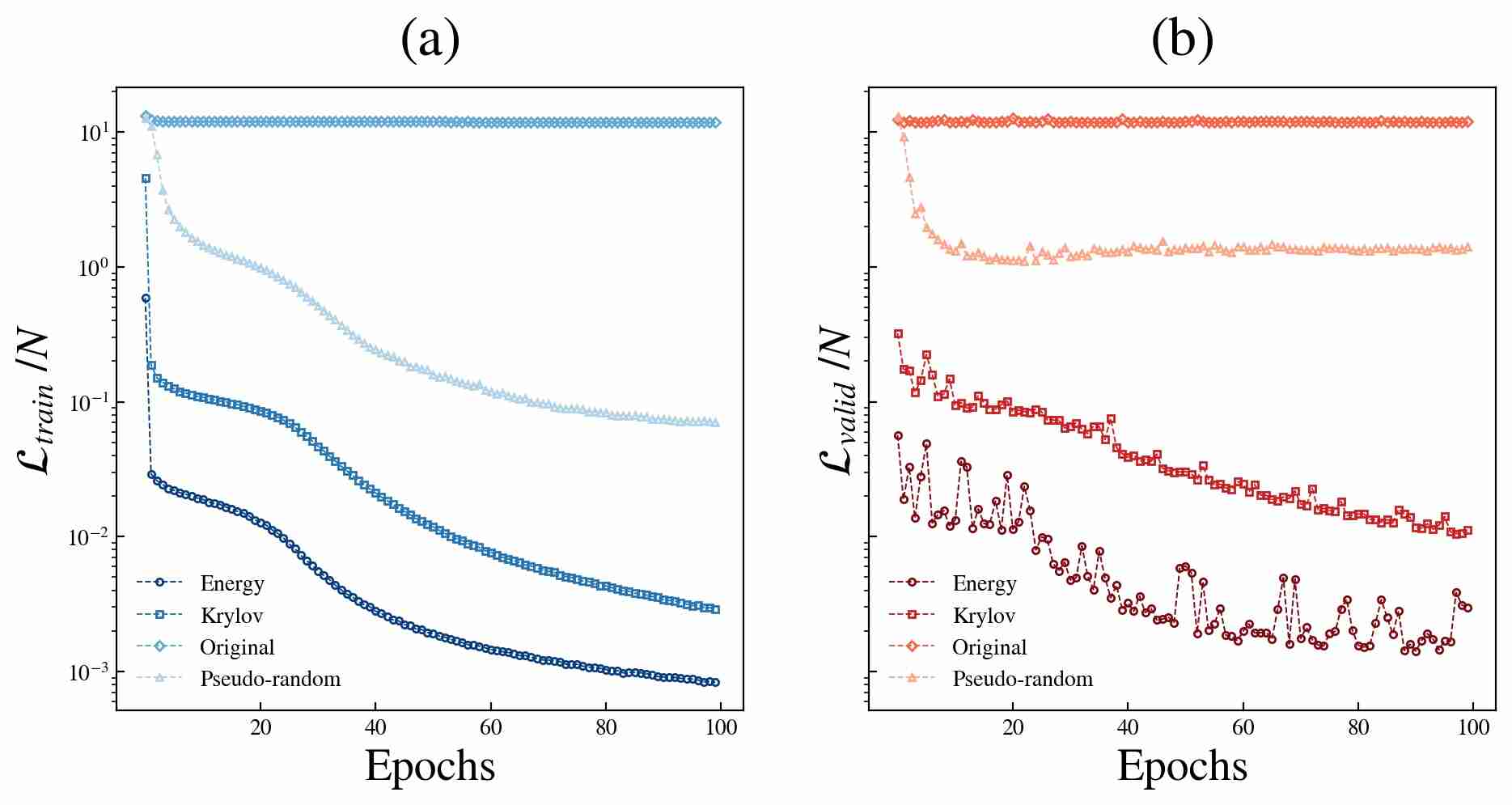}
  \caption{\small Loss function versus epoch for (a) training and (b) validation during training. 
  Circles, squares, diamonds, and triangles are for the energy, Krylov, original, and pseudo-random bases, respectively.
}
  \label{Fig_loss}
\end{figure}

Our model is trained using a supervised learning approach, where the algorithm learns to map inputs to their corresponding outputs based on labeled training data. Specifically, we employ a dataset of $N\,(=1024)$-dimensional state vectors paired with corresponding quantum complexity values. In this study, the entire dataset consists of $M=200$ samples of GUE, where each sample gives $3N=3072$ state vector data points with $3N$ denoting the total duration of our integer-valued time variable. Among them, 160 samples (\,or $160\times 3N$ data points\,) are used for training, while $20\times 3N$ data points are used for validation, and another $20\times 3N$ data points are used for testing. The loss function $\mathcal{L}$ used is mean squared error (MSE), 
\begin{align}\label{Loss}
  \mathcal{L}(W,b) = \frac{1}{\mathcal{N}} \sum_{I=1}^{\mathcal{N}} \left( C_I(W,b) - \hat{C}_I \right)^2,
\end{align}
where $C_I$ is the predicted quantum complexity value for the $I$-th input data, $\hat{C}_I$ is the true quantum complexity value, and batch size $\mathcal{N}$, the number of training examples used in one iteration of the model parameters ($W,b$) update, is set to 32\footnote{The batch size 
32 (taken as some power of two) is a common choice because it provides a balance between computational efficiency and model convergence. 
A smaller batch size (like 1) leads to noisy updates and slower convergence, while a larger batch size (equal to the entire dataset) makes the training process inefficient by updating too infrequently. 
The optimal batch size is usually determined through experimentation, balancing speed and accuracy.
}.
However, when calculating the validation loss, 
instead of $\mathcal{N}=32$,
we used the total number of data points in the validation dataset 
as the channel number {\it i.e.}
$\mathcal{N}=20\times3N=61,440$. Similarly, when calculating the test error, which is the root mean squared error (RMSE), we also use the total number of data points in the test dataset, so $\mathcal{N}=20\times3N=61,440$.
The loss function is minimized with respect to the model parameters $W$ and $b$ using the Adam optimizer \cite{kingma:2017adamCLP}, an adaptive learning rate optimization algorithm known for its efficiency and effectiveness in training DL models. The model is trained for a total of 100 epochs, where an epoch is one complete pass through the entire training dataset, since the training set has a total of 
$160\times 3N=491,520$ data points, and the batch size is 32, each epoch corresponds to 15,360 training iterations. In Figure~\ref{Fig_loss}, we plot loss functions against epochs for several different bases to examine the convergence behavior of the models during training. The convergence test is an important task in numerical calculations to illustrate the capability of our model and to detect potential problems during training. 
As can be seen in Figure~\ref{Fig_loss}, the stable convergence of loss functions with respect to epochs was obtained for the energy and Krylov bases. Perhaps due to the random nature of the original basis, our training algorithms 
get stuck immediately. For the pseudo-random basis, both training and validation processes still converge to a much larger value of the loss function compared to those of the energy and Krylov bases. Contrary to expectations, our results suggest that the choice of basis appears to be crucial in applying DL to the complexity.

\section{Results}\label{sec4}

\subsection{Basis dependence of DL performances}\label{sec4.1}

\begin{figure}[t] 
  \centering
  \includegraphics[width=16cm]{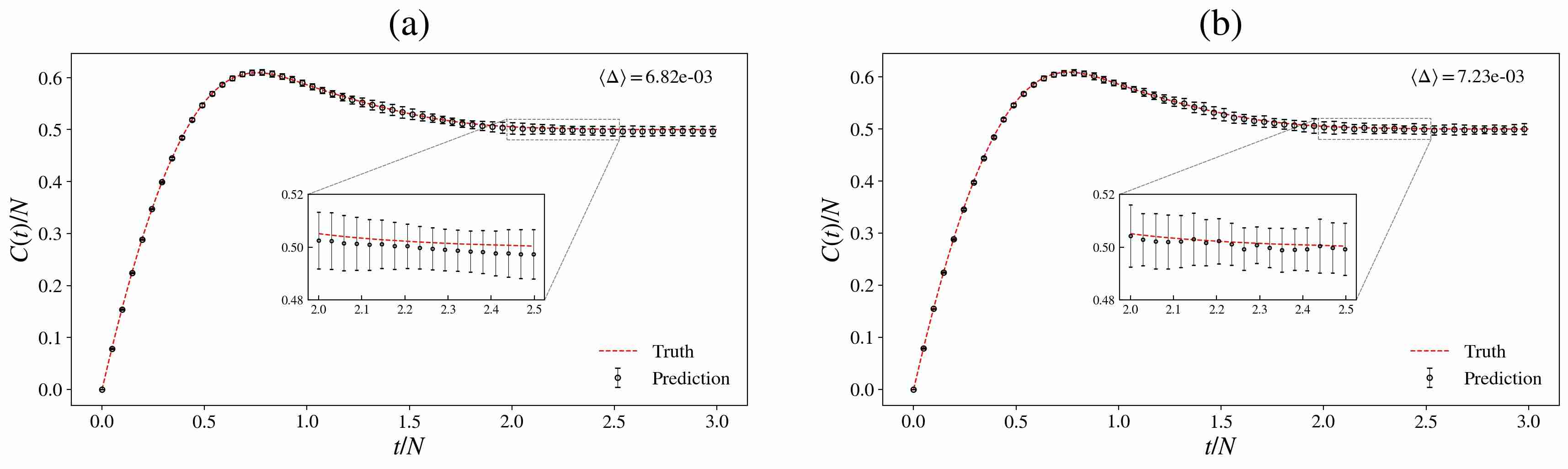}
  \caption{\small The comparison of $C(t)/N$ determined by CNN and truth values for (a) the energy and (b) the Krylov basis. The error bars denote the RMSE, $\Delta = \sqrt{\frac{1}{M_{test}}\sum_{i=1}^{M_{test}} (p_i/N - t_i/N)^2}$, where $p_i/N$ and $t_i/N$ are the $i$-th element of the predicted and truth values divided by $N$, respectively. Here we considered the TFD state with $N=1024$ for $\beta=0$. Note that $\langle \Delta \rangle$ is the time-averaged RMSE. The dotted lines are guides to the eye.}
  \label{Fig_EnKn}
\end{figure} 

\begin{figure}[t] 
  \centering
  \includegraphics[width=16cm]{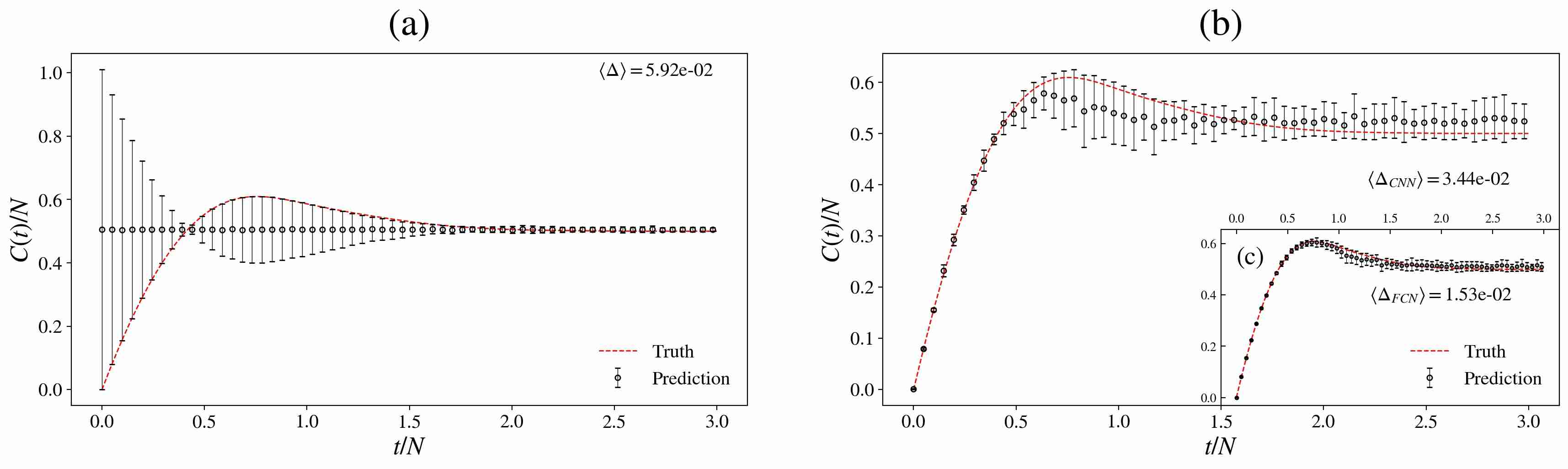}
  \caption{\small The comparison of $C(t)/N$ determined by CNN and truth values for (a) the original and (b) the pseudo-random basis. The inset (c) shows the results obtained by FCN.}
  \label{Fig_RnpRn}
\end{figure}  

Our primary interest in this paper is to demonstrate the validity and application of DL using CNN in quantum systems under specific conditions. Figures~\ref{Fig_EnKn} and~\ref{Fig_RnpRn}  illustrate the performances of our DL model across various basis sets.
Specifically, we compare the normalized Krylov spread complexity 
$C(t)/N$, 
as estimated by our CNN algorithms,
with the corresponding truth values as a function of time $t/N$. Remarkably, as shown in Figure~\ref{Fig_EnKn}, DL appears to perform well for the energy and Krylov bases. As can be seen in Figure~\ref{Fig_RnpRn}{\color{blue}(b)}, the choice of the pseudo-random basis also leads to a moderate overall performance of DL using CNN. As shown in the inset of Figure~\ref{Fig_RnpRn}{\color{blue}(b)}, further improvements have been made using FCN. However, the current DL model fails to describe the Krylov spread complexity for the original basis, as seen in Figure~\ref{Fig_RnpRn}{\color{blue}(a)}. Hence, the choice of basis is crucial in this case. At early times, our DL model, except for the original basis, performs extremely well in determining $C(t)/N$ until it reaches a peak. At late times, the DL performs reasonably well in the regions corresponding to the slope and plateau phases of complexity, despite larger sampling variations. It is important to note that the failure of the current DL model using CNN to predict the complexity in the original basis demands clarifications. While complexity, as defined in 
(\ref{Eq_Complexity}), is basis-independent, the performance of DL is clearly affected by the choice of basis. A particular comparison of the original and pseudo-random bases in the context of complexity could provide valuable insights into the underlying differences related to the nature of a basis set. No definite pattern or distinct feature from Figures~\ref{Fig_state}{\color{blue}(c)} and 
 {\color{blue}(d)} has been found. Clearly the success or failure of our DL model rests upon the slightly different circumstances, as illustrated in Figures~\ref{Fig_RnpRn}{\color{blue}(a)} and 
 {\color{blue}(b)}.

As explained in detail in Section \ref{sec2.1}, the state represented in the original basis for each sample random Hamiltonian becomes completely featureless. Due to the randomness of the basis transformations, our DL algorithms fail to extract the corresponding information on the spread complexity. In the case of the pseudo-random basis, we are simply adding a fixed noise, which is generally known to be effectively handled by DL algorithms. One may also attempt to include each sample Hamiltonian as input data for the DL, but with this inclusion, we find that the aforementioned features of basis dependence do not essentially change.

\subsection{Distinguishability for various $\beta$ states}\label{sec4.2}
\begin{figure}[t] 
  \centering
  \includegraphics[width=15.5cm]{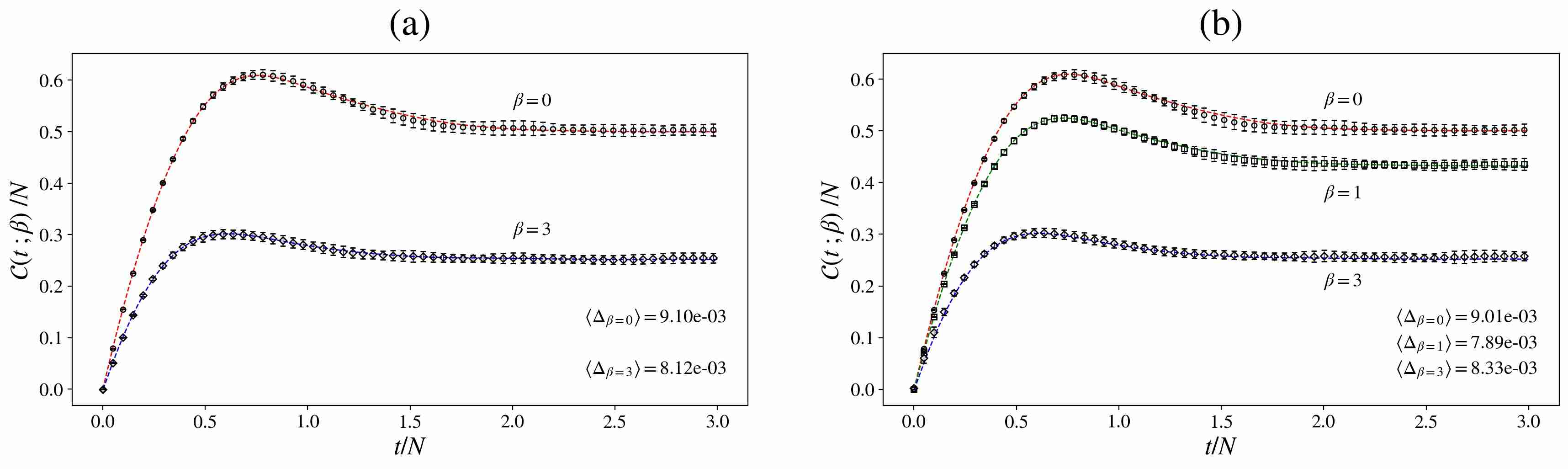}
  \caption{\small Temperature dependence of $C(t)/N$ distinguished by CNN and truth values for two (a) and three (b) different temperatures. Here we used the energy basis set. Circles, squares, and diamonds are for $\beta=0$, $\beta=1$, and $\beta=3$, respectively.}
  \label{Fig_beta}
\end{figure}

Although our DL model, with a suitable choice of basis sets, performs well in predicting complexity using CNN in the infinite-temperature limit, it remains unclear whether it will achieve similar efficiency and effectiveness in the finite-temperature case.
Our focus is to demonstrate that the DL model using CNN can distinguish samples 
with different temperatures.
In Figure~\ref{Fig_beta}, we show the temperature dependence of complexity $C(t)/N$ distinguished by DL 
using CNN for several different temperatures. 
The prediction has been made using random sequential samples at two and three temperatures,
as shown in Figures~\ref{Fig_beta}{\color{blue}(a)} and {\color{blue}(b)}, respectively. The energy basis was chosen due to its superior performance in previous benchmark tests. As expected, our results indicate that DL can differentiate key features in samples from multiple temperatures.

\subsection{System time variable 
and complexity}\label{sec4.3}

\begin{figure}[t] 
  \centering
  \includegraphics[width=15.5cm]{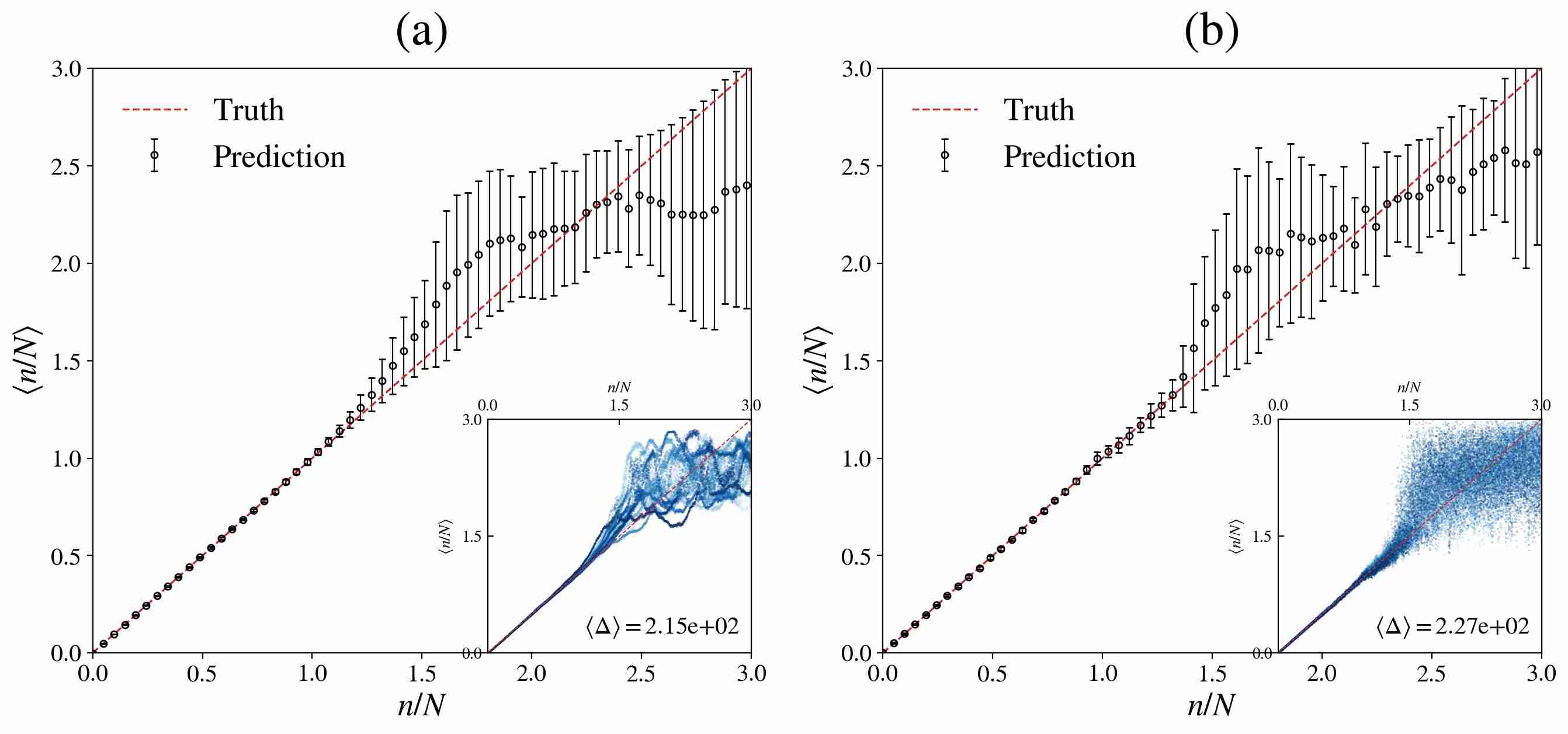}
  \caption{\small Average 
system time variable 
$\langle n/N \rangle$ estimated by CNN and the truth values for (a) the energy and (b) the Krylov basis. Insets show results in all samples predicted. The lines are guides to the eye.}
  \label{Fig_MLTime}
\end{figure}

Importantly, our findings suggest that the energy and Krylov basis sets are well-suited for training and predicting the complexity using DL models. Complexity, as a relevant physical quantity, is effectively captured by DL models, which can identify key features at specific times. It is intriguing to consider whether similar DL applications could be employed to evaluate other physical quantities. An immediate question arises: can the system time estimated by the DL model be regarded as a relevant quantity?
For this, let us introduce a unitary operator $U=e^{-iH}$, which evolves a state by a unit-interval of time. Then the integer-valued
time evolution 
is represented as $|\psi (n)\rangle = U^n\,|\psi (0)\rangle$ where the system time $n$ corresponds to
the number of $U$ operations with $n=0,1,2,\cdots,3N$.
Figure~\ref{Fig_MLTime} shows the average system time variable 
$\langle n/N \rangle$  determined by DL using CNN, compared to the truth values
for energy and Krylov bases. (This number $n$ should not be confused with the quantum-number 
index $n$ in Figure~\ref{Fig_state}.) Both bases were used because of their previously demonstrated performance. Although the DL model appears to perform well for small time regions ($n/N \lesssim 1.5$), significantly large fluctuations in the average system time $\langle n/N \rangle$ are observed for $n/N\gtrsim 1.5$. 
Therefore, our results suggest that the integer-valued system time is irrelevant for characterizing states.

\section{Conclusions}\label{sec5}

In this note, we study how a machine based on DL algorithms learns the Krylov spread complexity from patterns of quantum states in systems with $N\times N$ random Hamiltonians drawn from the GUE of matrices. Taking the initial state as the TFD state, we found that the DL algorithm performs well for states at any time, including the plateau region where the states become almost featureless and random.
The TFD initial state in our DL setup could be generalized to nonthermal states that span a broader region of the state space.

We also explored other variables, such as the system time variable of a quantum state, as a replacement for the Krylov spread complexity. However, in this case, we found that the DL algorithm fails in the late-time region, with the estimated learning error becoming very large. This suggests that the Krylov spread complexity is effective in characterizing the structure of quantum states.

The deep learning performance of the Krylov operator complexity \cite{Parker:2018yvk} is of interest, given that its state space structure differs from that of the Krylov spread complexity. Our investigation of the Krylov spread complexity in this note may be extended to other systems, such as random matrix theories with
other ensembles, field theories \cite{Jefferson:2017sdb,Dymarsky:2021bjq,Avdoshkin:2022xuw}, non-chaotic systems \cite{Rabinovici:2022beu}, and so on. It will be particularly interesting to see whether deep learning algorithms can effectively distinguish different random matrix ensembles or between chaotic and non-chaotic systems.
Here, we use $N=1024$, but studying the learning efficiency and the associated scaling behavior as
$N$ changes may also be of interest. However, this would involve 
computational challenges and warrants further investigation.

\subsection*{Acknowledgement}
D.B. was supported by the 2024 Research Fund of the University of Seoul.
This work was also supported by the UBAI computing resources at the University of Seoul.





  \setcounter{equation}{0}

%


\begin{thebibliography}{99}\label{bibg}



\bibitem{Susskind:2014rva}
L.~Susskind,
``Computational Complexity and Black Hole Horizons,''
Fortsch. Phys. \textbf{64}, 24-43 (2016)
[arXiv:1403.5695 [hep-th]].

\bibitem{Brown:2015lvg}
A.~R.~Brown, D.~A.~Roberts, L.~Susskind, B.~Swingle and Y.~Zhao,
``Complexity, action, and black holes,''
Phys. Rev. D \textbf{93}, no.8, 086006 (2016)
[arXiv:1512.04993 [hep-th]].



\bibitem{Brown:2015bva}
A.~R.~Brown, D.~A.~Roberts, L.~Susskind, B.~Swingle and Y.~Zhao,
``Holographic Complexity Equals Bulk Action?,''
Phys. Rev. Lett. \textbf{116}, no.19, 191301 (2016)
[arXiv:1509.07876 [hep-th]].


\bibitem{Belin:2021bga}
A.~Belin, R.~C.~Myers, S.~M.~Ruan, G.~S\'arosi and A.~J.~Speranza,
``Does Complexity Equal Anything?,''
Phys. Rev. Lett. \textbf{128}, no.8, 081602 (2022)
[arXiv:2111.02429 [hep-th]].


\bibitem{Iliesiu:2021ari}
L.~V.~Iliesiu, M.~Mezei and G.~S\'arosi,
``The volume of the black hole interior at late times,''
JHEP \textbf{07}, 073 (2022)
[arXiv:2107.06286 [hep-th]].




\bibitem{Caputa:2021sib}
P.~Caputa, J.~M.~Magan and D.~Patramanis,
``Geometry of Krylov complexity,''
Phys. Rev. Res. \textbf{4}, no.1, 013041 (2022)
[arXiv:2109.03824 [hep-th]].

\bibitem{Brown:2021euk}
A.~R.~Brown,
``A quantum complexity lower bound from differential geometry,''
Nature Phys. \textbf{19}, no.3, 401-406 (2023)
[arXiv:2112.05724 [hep-th]].



\bibitem{Rabinovici:2020ryf}
E.~Rabinovici, A.~S\'anchez-Garrido, R.~Shir and J.~Sonner,
``Operator complexity: a journey to the edge of Krylov space,''
JHEP \textbf{06}, 062 (2021)
[arXiv:2009.01862 [hep-th]].

\bibitem{Zhai:2024tkz}
K.~H.~Zhai, L.~H.~Liu and H.~Q.~Zhang,
``The generalized CV conjecture of Krylov complexity,''
[arXiv:2412.08925 [hep-th]].

\bibitem{Balasubramanian:2022tpr}
V.~Balasubramanian, P.~Caputa, J.~M.~Magan and Q.~Wu,
``Quantum chaos and the complexity of spread of states,''
Phys. Rev. D \textbf{106}, no.4, 046007 (2022)
[arXiv:2202.06957 [hep-th]].

\bibitem{Nandy:2024evd}
P.~Nandy, A.~S.~Matsoukas-Roubeas, P.~Mart{\'\i}nez-Azcona, A.~Dymarsky and A.~del Campo,
``Quantum dynamics in Krylov space: Methods and applications,''
Phys. Rept. \textbf{1125-1128}, 1-82 (2025)
[arXiv:2405.09628 [quant-ph]].

\bibitem{Parker:2018yvk}
D.~E.~Parker, X.~Cao, A.~Avdoshkin, T.~Scaffidi and E.~Altman,
``A Universal Operator Growth Hypothesis,''
Phys. Rev. X \textbf{9}, no.4, 041017 (2019)
[arXiv:1812.08657 [cond-mat.stat-mech]].





\bibitem{Erdmenger:2023wjg}
J.~Erdmenger, S.~K.~Jian and Z.~Y.~Xian,
``Universal chaotic dynamics from Krylov space,''
JHEP \textbf{08}, 176 (2023)
[arXiv:2303.12151 [hep-th]].





\bibitem{Cotler:2017jue}
J.~Cotler, N.~Hunter-Jones, J.~Liu and B.~Yoshida,
``Chaos, Complexity, and Random Matrices,''
JHEP \textbf{11} (2017), 048
[arXiv:1706.05400 [hep-th]].

\bibitem{Hornedal:2022pkc}
N.~H{\"o}rnedal, N.~Carabba, A.~S.~Matsoukas-Roubeas and A.~del Campo,
``Ultimate Speed Limits to the Growth of Operator Complexity,''
Commun. Phys. \textbf{5}, 207 (2022)
[arXiv:2202.05006 [quant-ph]].

\bibitem{Brandao:2019sgy}
F.~G.~S.~L.~Brand\~ao, W.~Chemissany, N.~Hunter-Jones, R.~Kueng and J.~Preskill,
``Models of Quantum Complexity Growth,''
PRX Quantum \textbf{2}, no.3, 030316 (2021)
[arXiv:1912.04297 [hep-th]].


\bibitem{Haferkamp:2021uxo}
J.~Haferkamp, P.~Faist, N.~B.~T.~Kothakonda, J.~Eisert and N.~Y.~Halpern,
``Linear growth of quantum circuit complexity,''
Nature Phys. \textbf{18}, no.5, 528-532 (2022)
[arXiv:2106.05305 [quant-ph]].


\bibitem{Guhr:1997ve}
T.~Guhr, A.~Muller-Groeling and H.~A.~Weidenmuller,
``Random matrix theories in quantum physics: Common concepts,''
Phys. Rept. \textbf{299} (1998), 189-425
[arXiv:cond-mat/9707301 [cond-mat]].

\bibitem{Bhattacharjee:2024yxj}
B.~Bhattacharjee and P.~Nandy,
``Krylov fractality and complexity in generic random matrix ensembles,''
Phys. Rev. B \textbf{111}, no.6, L060202 (2025)
[arXiv:2407.07399 [quant-ph]].

\bibitem{Cotler:2016fpe}
J.~S.~Cotler, G.~Gur-Ari, M.~Hanada, J.~Polchinski, P.~Saad, S.~H.~Shenker, D.~Stanford, A.~Streicher and M.~Tezuka,
``Black Holes and Random Matrices,''
JHEP \textbf{05}, 118 (2017)
[erratum: JHEP \textbf{09}, 002 (2018)]
[arXiv:1611.04650 [hep-th]].

\bibitem{Maldacena:2001kr}
J.~M.~Maldacena,
``Eternal black holes in anti-de Sitter,''
JHEP \textbf{04}, 021 (2003)
[arXiv:hep-th/0106112 [hep-th]].


\bibitem{LeCun:2015pmr}
Y.~LeCun, Y.~Bengio and G.~Hinton,
``Deep learning,''
Nature \textbf{521}, 436-444 (2015).

\bibitem{Lecun:1998iof}
Y.~Lecun, L.~Bottou, Y.~Bengio and P.~Haffner,
``Gradient-based learning applied to document recognition'', Proceedings of the IEEE, vol. 86, no. 11, pp. 2278-2324, Nov. 1998.



\bibitem{Glorot:2010AISTATS}
Xavier Glorot Yoshua Bengio,
``Understanding the difficulty of training deep feedforward neural networks'',
Proceedings of the 13th International Conference on Artificial Intelligence and Statistics (AISTATS) 2010.


\bibitem{Lanczos:1950zz}
C.~Lanczos,
``An iteration method for the solution of the eigenvalue problem of linear differential and integral operators,''
J. Res. Natl. Bur. Stand. B \textbf{45} (1950), 255-282.




\bibitem{kingma:2017adamCLP}
Diederik P. Kingma and Jimmy Ba,
``Adam: A Method for Stochastic Optimization'',
Conference paper at 3rd International Conference for Learning Representations, San Diego, 2015.

\bibitem{Jefferson:2017sdb}
R.~Jefferson and R.~C.~Myers,
``Circuit complexity in quantum field theory,''
JHEP \textbf{10}, 107 (2017)
[arXiv:1707.08570 [hep-th]].



\bibitem{Dymarsky:2021bjq}
A.~Dymarsky and M.~Smolkin,
``Krylov complexity in conformal field theory,''
Phys. Rev. D \textbf{104}, no.8, L081702 (2021)
[arXiv:2104.09514 [hep-th]].


\bibitem{Avdoshkin:2022xuw}
A.~Avdoshkin, A.~Dymarsky and M.~Smolkin,
``Krylov complexity in quantum field theory, and beyond,''
JHEP \textbf{06}, 066 (2024)
[arXiv:2212.14429 [hep-th]].


\bibitem{Rabinovici:2022beu}
E.~Rabinovici, A.~S\'anchez-Garrido, R.~Shir and J.~Sonner,
``Krylov complexity from integrability to chaos,''
JHEP \textbf{07}, 151 (2022)
[arXiv:2207.07701 [hep-th]].









\end{thebibliography}
\end{document}